\documentclass[12pt]{article}

\usepackage{latexsym}

\usepackage{graphics}
\usepackage{graphicx}

\begin{document}


\title{Sequences of dipole black rings and Kaluza-Klein bubbles}

\author{
     Stoytcho S. Yazadjiev \thanks{E-mail: yazad@phys.uni-sofia.bg}, Petia G. Nedkova\thanks{E-mail:pnedkova@phys.uni-sofia.bg}\\
{\footnotesize  Department of Theoretical Physics,
                Faculty of Physics, Sofia University,}\\
{\footnotesize  5 James Bourchier Boulevard, Sofia~1164, Bulgaria }\\
}

\date{}

\maketitle

\begin{abstract}
We construct new exact solutions to 5D Einstein-Maxwell equations describing sequences of Kaluza-Klein bubbles and dipole black rings.
The solutions are generated by 2-soliton transformations from vacuum black ring - bubble sequences. The properties of the solutions are
investigated. We also derive the Smarr-like relations and the mass and tension first laws in the general case for such configurations of Kaluza-Klein bubbles and dipole black rings. The novel moment is the appearance of the magnetic flux in the Smarr-like relations and the first laws.

\end{abstract}


\sloppy

\section{Introduction}

Higher dimensional gravity and especially black holes in higher dimensional spacetimes became established and very active area of research.
Higher dimensional black holes exhibit very interesting properties and features some of which are completely absent in four dimensions.
Of particular interest are the black holes in spacetimes with compact extra dimensions (with circle topology in most cases) known as Kaluza-Klein black holes. More precisely, a five dimensional  spacetime ${\cal M}$ (the case we consider here) is called Kaluza-Klein
spacetime if  it is asymptotically ${\cal M}^{4}\times S^1$ where ${\cal M}^{4}$ is the 4-dimensional Minkowski spacetime. Very recently the uniqueness theorem for vacuum Kaluza-Klein black holes was established in \cite{HY}.
This theorem gives complete  classification of the  possible horizon topologies and classification of the black solutions on the basis of the so-called
interval (rod)\footnote{With regard to the general concept of rod structure we refer the reader to \cite{ER2}. More precise mathematical definition of the  rod structure (the so called interval structure) can be found in \cite{HY}.} structure. Some exact Kaluza-Klein black hole solutions\footnote{
It is worth mentioning that there are interesting Kaluza-Klein black hole solutions which have an  asymptotic different from the standard one
${\cal M}^{4}\times S^1$.  For explicit examples we refer the reader to \cite{SSW}-\cite{Y_squashed}. } have also been constructed \cite{ER2}-\cite{KY}. Among them are the solutions describing
sequences of static vacuum black holes and bubbles \cite{EH}-\cite{EHO} which are of particular interest in the context of the present paper.
We also refer the reader to the review article \cite{HNO} where the Kaluza-Klein black holes are considered from different perspectives.
Nevertheless, the known exact Kaluza-Klein black hole solutions are far from being exhaustive.
The thermodynamics of Kaluza-Klein black holes   in the presence of Maxwell field along the compact dimension exhibits novel and interesting features \cite{YazadjievNedkova}. For example, new terms related to the magnetic flux  appear in the  Smarr-like relations and the first laws.

In the present paper we are dealing  with sequences of 5D  Kaluza-Klein bubbles and black rings in the presence of self-gravitating Maxwell field.
While in the static vacuum case the construction of such configurations can be done in relatively simple way by solving linear equations, in the presence of self-gravitating Maxwell field, the construction of such configurations  is much more difficult since we are forced to solve nonlinear
equations.
In this paper we construct new exact solutions to the 5D Einstein-Maxwell gravity describing sequences of dipole black rings and
Kaluza-Klein bubbles. The solutions are generated by 2-soliton transformations from vacuum black ring - bubble configurations. The
basic physical quantities characterizing the new solutions are computed. We also derive the Smarr-like relations and the mass and tension first
laws for such configurations of dipole black rings and Kaluza-Klein bubbles in the general case.

\section{Solution generating method and the exact \\solutions}

\subsection{Solution generating method}

In five dimensions the Einstein-Maxwell  equations read
\begin{eqnarray}\label{EMFE}
&&R_{\mu\nu} = {1\over 2} \left(F_{\mu\lambda}F_{\nu}^{\,\lambda}
 - {1\over 6} F_{\sigma\lambda}F^{\sigma\lambda} g_{\mu\nu}\right),  \\
&&\nabla_{\mu} F^{\mu\nu} = \nabla_{[\mu}F_{\nu\lambda ]} =0  \nonumber.
\end{eqnarray}

In this paper  we consider 5D EM  gravity in  spacetimes with the symmetry group $R~\times U(1)^2$ generated by the commuting Killing fields
$\xi$, $\zeta$  and $\eta$. Here $\xi$ is the asymptotically timelike Killing field and  $\zeta$  and $\eta$ are the axial Killing fields, respectively. The Killing field $\eta$ will be associated with the compact dimension.
We also assume
that all the Killing fields  are  hypersurface orthogonal.
In this case, using adapted  coordinates in which $\xi=\partial/\partial {t}$, $\zeta=\partial/\partial \psi$ and $\eta=\partial/\partial \phi$,
the 5D spacetime metric can be written in the form
\begin{eqnarray}
ds^2= - e^{2\chi-u}dt^2 + e^{-2\chi-u}\rho^2 d\psi^2 + e^{-2\chi-u} e^{2\Gamma}\left(d\rho^2 + dz^2 \right) +
e^{2u}d\phi^2
\end{eqnarray}
where all the metric functions depend on the canonical coordinates $\rho$ and $z$ only.

For the electromagnetic field we impose the following conditions

\begin{eqnarray}
&&{\cal L}_{\xi}F= {\cal L}_{\zeta}F= {\cal L}_{\eta}F=0, \\
&&i_{\xi}F=i_{\zeta}F=i_{\eta} \star F=0 ,\nonumber
\end{eqnarray}
where $\star$ is the Hodge dual,  ${\cal L}_{X}$ denotes the Lie derivative along the vector field $X$ and $i_X$ is the interior product of the vector field $X$ with an arbitrary form. From a local point of view these conditions mean that the gauge potential\footnote{In the presence of dipole (magnetic) charges the gauge potential is not globally well-defined.} $A$ has the local form
$A=A_{\phi}d\phi$.

The 1-form $i_{\eta}F$ is invariant under the spacetime symmetries  and therefore  can be considered as 1-form on the factor space $\hat {\cal M}={\cal M}/R\times U(1)^2$. Since the factor space is simply connected \cite{HY} and $i_{\eta}F$ is closed ($di_{\eta}F=0$) there exists a globally well-defined potential $\lambda$ such that

\begin{eqnarray}\label{potential_lambda}
i_{\eta}F=-d\lambda.
\end{eqnarray}

It is worth mentioning that locally we have $\lambda=A_{\phi}$ up to a constant.

Further we introduce the complex Ernst potential ${\cal E}$  defined by
\begin{eqnarray}
&&{\cal E}= e^{u} + {i\over \sqrt{3}} \lambda .
\end{eqnarray}

With the help of the Ernst potentials the dimensionally reduced 5D Einstein-Maxwell equations can be written in the following form
\begin{eqnarray}
&&\left({\cal E} + {\cal E}^{*}\right)\left(\partial^2_{\rho} {\cal E} + \rho^{-1} \partial_{\rho}{\cal E}
+ \partial^2_{z} {\cal E}\right)= 2 \left(\partial_{\rho}{\cal E}\partial_{\rho}{\cal E}
+ \partial_{z}{\cal E}\partial_{z}{\cal E} \right), \nonumber
\end{eqnarray}
\begin{eqnarray} \label{REDUCEDEMEQUATIONS}
&&\partial^2_{\rho} \chi + \rho^{-1} \partial_{\rho}\chi
+ \partial^2_{z} \chi=0,
\end{eqnarray}
\begin{eqnarray}
&&\rho^{-1}\partial_{\rho}\Gamma=
\left(\partial_{\rho}\chi\right)^2 - \left(\partial_{z}\chi \right)^2
+ {3\over \left({\cal E} + {\cal E}^{*} \right)^2}
\left(\partial_{\rho}{\cal E}\partial_{\rho}{\cal E}^{*} - \partial_{z}{\cal E}\partial_{z}{\cal E}^{*} \right),\nonumber \\
\nonumber \\
 &&\rho^{-1}\partial_{z}\Gamma= 2\partial_{\rho}\chi \partial_{z}\chi
+ {6\over \left({\cal E} + {\cal E}^{*} \right)^2} \partial_{\rho}{\cal E} \partial_{z}{\cal E}^{*}.\nonumber
\end{eqnarray}

The consistency conditions for the last two equations in  (\ref{REDUCEDEMEQUATIONS}), i.e. the equations for the metric function $\Gamma$,
are guaranteed by the first two  equations in (\ref{REDUCEDEMEQUATIONS}).

In this way we reduced the problem of solving the 5D EM equations to two effective 4D problems, i.e. two Ernst equations.
The central and most difficult task is to solve the nonlinear Ernst equation. Here we will not discuss in detail the
methods for solving the Ernst equation. Instead we shall present the working formulas we need. Details can be found in \cite{Yazadjiev}.

Let us consider  a solution to the vacuum 5D Einstein equations

\begin{eqnarray}\label{seedsolution}
ds_{E}^2= g^{E}_{00}dt^2 + g^{E}_{\psi\psi}d\psi^2 + g^{E}_{\rho\rho}(d\rho^2 + dz^2) + g^{E}_{\phi\phi}d\phi^2
\end{eqnarray}
with metric function $g^{E}_{\phi\phi}$ given by
\begin{eqnarray}
g^{E}_{\phi\phi}=e^{2u_{0}}=\prod_{i=1}^{N}\left(e^{2{\tilde U}_{\nu_i}}\right)^{\epsilon_i},
\end{eqnarray}
where
\begin{eqnarray}
e^{2{\tilde U}_{\nu_i}}= R_{\nu_i} + \zeta_{\nu_i}=\sqrt{\rho^2 + \zeta^2_{\nu_i}} + \zeta_{\nu_i} =\sqrt{\rho^2 + (z-\nu_i)^2} + (z-\nu_i)
\end{eqnarray}
and $\nu_i$ and $\epsilon_i$ are constants.

The 2-soliton transformation generates the following solution to the 5D Einstein-Maxwell equations from the vacuum solution (\ref{seedsolution})

\begin{eqnarray}\label{EMSOLUTION}
&&ds^2= {g^{E}_{00}\over W} dt^2 + {g^{E}_{\psi\psi}\over W}d\psi^2 + {{\cal Y}^3\over W}g^{E}_{\rho\rho}(d\rho^2 + dz^2) + W^2g^{E}_{\phi\phi}d\phi^2, \nonumber
 \\ \\
&&\lambda= 4\sqrt{3}\Delta k e^{u_0} {[R_{k_1}a(1+b^2) + R_{k_2}b(1+a^2)]\over W_2} + \lambda_{0}, \nonumber
\end{eqnarray}
where $k_1$, $k_2$ ($\Delta k=k_1-k_2$) and $\lambda_{0}$ are constants. Without loss of generality we put $\lambda_{0}=0.$ The functions included in (\ref{EMSOLUTION}) are presented below. The functions $a$ and $b$ are given by

\begin{eqnarray}
a= \alpha \prod^{N}_{i=1}\left({e^{2U_{k_1}} + e^{2{\tilde U}_{\nu_i}}\over  e^{{\tilde U}_{\nu_i}}} \right)^{\epsilon_i}, \nonumber \\ \\
b=\beta \prod^{N}_{i=1}\left({e^{2U_{k_2}} + e^{2{\tilde U}_{\nu_i}}\over  e^{{\tilde U}_{\nu_i}}} \right)^{-\epsilon_i},\nonumber
\end{eqnarray}
where $\alpha$ and $\beta$ are constants and

\begin{eqnarray}
e^{2U_{k_i}}=R_{k_i} -\zeta_{k_i}=\sqrt{\rho^2 + (z-k_i)^2} -(z-k_i).
\end{eqnarray}

The function $W$ is presented in the form

\begin{eqnarray}
W={W_1\over W_2}
\end{eqnarray}
where

\begin{eqnarray}\label{functionsW}
W_1&=&\left[(R_{k_1}+R_{k_2})^2-(\Delta k)^2\right] (1+ a b)^2  + \left[(R_{k_1}-R_{k_2})^2-(\Delta k)^2\right](a-b)^2
 ,\nonumber \\  \\
W_2&=&\left[(R_{k_1}+R_{k_2}+\Delta k)+(R_{k_1}+R_{k_2}-\Delta k)a b \right]^2 \nonumber \\
 && +\left[(R_{k_1}-R_{k_2}-\Delta k)a - (R_{k_1}-R_{k_2}+\Delta k)b \right]^2 .  \nonumber
\end{eqnarray}

For the function ${\cal Y}$ we have
\begin{eqnarray}
{\cal Y} = {\cal Y}_{0}{W_1\over (R_{k_1}+R_{k_2})^2 - (\Delta k)^2} e^{2h}
\end{eqnarray}
where ${\cal Y}_0$ is a constant and
\begin{eqnarray}
h= \gamma_{k_1,k_1} - 2\gamma_{k_1,k_2} + \gamma_{k_2,k_2} + \sum^{N}_{i}\epsilon_{i}\left(\gamma_{k_1,\nu_i}-\gamma_{k_2,\nu_i}\right),
\end{eqnarray}

\begin{eqnarray}
\gamma_{k,l}={1\over 2}{\tilde U}_{k} + {1\over 2}{\tilde U}_{l} - {1\over 4}\ln[R_kR_l + (z-k)(z-l) + \rho^2].
\end{eqnarray}

Before closing this subsection we should explain the following. The 2-soliton transformation involves two parameters $k_1$ and $k_2$
but one of them can be absorbed by a shift  $z\to z + constant$ under which the  "electromagnetic part" of the field equations is invariant.
We prefer to keep using  the two parameters  $k_1$ and $k_2$ but those readers who feel unconformable with these parameters
may think that parameters $k_1$ and $k_2$ are appropriate functions only of $\Delta k$
and examples will be given below (see eq.(\ref{choice_k1_k2})).

\subsection{The exact solutions}

\begin{figure}
\begin{center}
      \includegraphics[width=12cm]{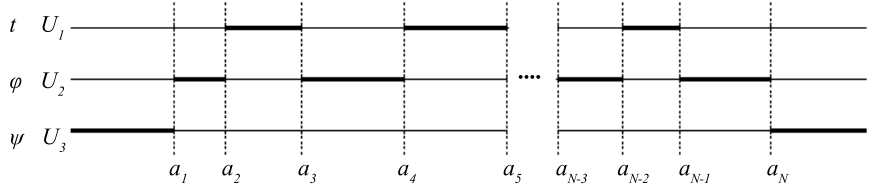}

\caption{Rod structure of the bubble - black ring sequence} \label{rodstrn1}
       \end{center}
\end{figure}

In order to generate the solutions to  the 5D Einstein-Maxwell equations describing sequence of dipole black rings and Kaluza-Klein (KK) bubbles  we take as a seed the vacuum solution describing neutral black rings\footnote{It should be noted that the topology of the black rings is $S^2\times S^1$ where
$S^1$ is not topologically supported, i.e. $S^1$ is associated with the orbits of $\zeta$ and is not the Kaluza-Klein circle. This can be seen from the rod diagram (\ref{rodstrn1}).} on KK bubbles  constructed in \cite{EHO}. More precisely this solution describes a sequence of $q$ black rings and $p=q+1$ KK bubbles in the form
\\

{\bf bubble} -- {\bf black ring} -- {\bf bubble} -- {\bf black ring} -- ... --{\bf bubble} -- {\bf black ring} -- {\bf bubble}
\\

with rod structure shown on figure (\ref{rodstrn1}). There are $q$ finite rods $[a_2,a_3], [a_4, a_5],....,[a_{N-2},a_{N-1}]$ corresponding to the black ring horizons and $p$ finite rods $[a_1,a_2], [a_3,a_4],...,[a_{N-1}, a_{N}]$ corresponding to the KK bubbles, i.e. the axes of the Killing vector ${\partial/\partial \phi}$. The semi-infinite rods $[-\infty, a_1]$ and $[a_N, +\infty]$ describe the axes of the Killing vector ${\partial/\partial \psi}$. The number $N$ is even and is given by $N=2(q+1)$. For simplicity  we will confine here to the particular case of one black ring surrounded by two KK bubbles, i.e. we will consider $q = 1$ and $p = 2$. The generalization to sequences consisting of arbitrary number of bubbles and black rings is straightforward and it is presented in Appendix A. The explicit analytical form of the solution in the specified case is the following

\begin{eqnarray}
&&g^{E}_{00}= - \frac{(R_{a_2}-\zeta_{a_2})}{(R_{a_3}-\zeta_{a_3})}= - \prod^{3}_{i=2} \left(e^{2U_{a_i}}\right)^{(-1)^{i}}, \\ \nonumber \\
&&g^{E}_{\phi\phi}=  \frac{(R_{a_1}-\zeta_{a_1})(R_{a_3}-\zeta_{a_3})}{(R_{a_2}-\zeta_{a_2})(R_{a_4}-\zeta_{a_4})}=\prod_{i=1}^{4} \left(e^{2U_{a_i}}\right)^{(-1)^{i+1}}, \\ \nonumber \\
&&g^{E}_{\psi\psi}= (R_{a_1} + \zeta_{a_1})(R_{a_4}- \zeta_{a_4}), \\ \nonumber\\
&&g^{E}_{\rho\rho}= \frac{Y_{14}Y_{23}}{4 R_{a_1} R_{a_2} R_{a_3} R_{a_4}}\sqrt{\frac{Y_{12}Y_{34}}{Y_{24}Y_{13}}}\frac{R_{a_4}-\zeta_{a_4}}{ R_{a_1}-\zeta_{a_1}},
\end{eqnarray}
where

\begin{eqnarray}
Y_{ij}= R_{a_i}R_{a_j} +\zeta_{a_i}\zeta_{a_j} + \rho^2 .
\end{eqnarray}

Taking into account that
\begin{eqnarray}
e^{2U_{a_i}}=\rho^2 e^{-2{\tilde U}_{a_i}}
\end{eqnarray}
and $\sum^{N}_{i=1} (-1)^{i+1}=0$ for even $N$,  it is not difficult to see that

\begin{eqnarray}
g^{E}_{\phi\phi}=\prod_{i=1}^{4} \left(e^{2{\tilde U}_{a_i}}\right)^{(-1)^{i}}.
\end{eqnarray}

For the seed solution under consideration we find

\begin{eqnarray}
&&a=\alpha \prod^{4}_{i=1} \left( {e^{2U_{k_1}} + e^{2{\tilde U}_{a_i}}}\over e^{{\tilde U}_{a_i}}\right)^{(-1)^{i}}, \\ \nonumber\\
&&b= \beta \prod^{4}_{i=1} \left( {e^{2U_{k_2}} + e^{2{\tilde U}_{a_i}}}\over e^{{\tilde U}_{a_i}}\right)^{(-1)^{i+1}}.
\end{eqnarray}

\section{Analysis  of the solution}

The investigation of the solution shows that the functions $W$, ${\cal Y }$ and $\lambda$, under some conditions discussed below,
are regular everywhere for the following ordering of the parameters

\begin{eqnarray}
a_{2m-1} <k_2 <k_1 <a_{2m}
\end{eqnarray}
where $m=1,2$. In other words the parameters $k_1$ and $k_2$ must lie on any of the bubble rods.

A convenient choice for the parameters
$k_1$ and $k_2$ is the following

\begin{eqnarray}\label{choice_k1_k2}
k_1= {a_{2m}+ a_{2m-1}\over 2} + {1\over 2} \Delta k,\\
k_2= {a_{2m}+ a_{2m-1}\over 2} - {1\over 2} \Delta k.\nonumber
\end{eqnarray}

There are potential singularities in the functions $W$ and ${\cal Y}$ for $z=k_1$ and $z=k_2$. In order to eliminate these potential singularities we must impose

\begin{eqnarray}
&&\alpha^2 = \prod^{2m-1}_{i=1}(k_1-a_i)^{(-1)^{i+1}} \prod^{4}_{j=2m}(a_j-k_1)^{(-1)^{j+1}} ,   \\ \nonumber   \\
&&\beta^2 = \prod^{2m-1}_{i=1}(k_2-a_i)^{(-1)^{i}} \prod^{4}_{j=2m}(a_j-k_2)^{(-1)^{j}} .
\end{eqnarray}

\subsection{Asymptotics}

In order to study the asymptotic behavior of the solution we introduce the asymptotic coordinates
$r$ and $\theta$ defined by

\begin{eqnarray}
&&\rho=r\sin\theta,  \,\, z=r\cos\theta. \nonumber
\end{eqnarray}

Then in the asymptotic limit we find

\begin{eqnarray}
&&g^{E}_{00} \approx -1 + {1\over r} \sum^{q=1}_{s=1} (a_{2s+1}- a_{2s}) = -1 + \frac{c^E_t}{r},\\  \nonumber \\
&&g^{E}_{\psi\psi}\approx r^2\sin^2\theta ,\\\nonumber \\
&&g^{E}_{\phi\phi} \approx 1 - {1\over r} \sum^{p=2}_{s=1}(a_{2s} - a_{2s-1}) = 1 + \frac{c^E_\phi}{r},\\ \nonumber\\
&&g^{E}_{\rho\rho} \approx 1, \\\nonumber \\
&& W \approx  1 - {1- \alpha\beta\over 1 + \alpha\beta}{\Delta k\over r}, \\ \nonumber\\
&& {\cal Y} \approx  {\cal Y}_{0}(1 + \alpha\beta)^2 ,\\\nonumber\\
&& \lambda \approx \sqrt{3}{ \alpha + \beta \over  1 + \alpha\beta}{\Delta k\over r}.
\end{eqnarray}

In order for our solution to be asymptotically Kaluza--Klein we must impose

\begin{eqnarray}
{\cal Y}_{0}= {1\over (1 + \alpha\beta)^2}.
\end{eqnarray}

\subsection{Balance conditions}

Let us first consider the semi-infinite rods corresponding to the axes of the Killing vector ${\partial/\partial \psi}$. The regularity conditions
then give the following period of $\psi$

\begin{eqnarray}
\Delta \psi = 2\pi \lim_{\rho\to 0}\sqrt{{\rho^2 g_{\rho\rho}\over g_{\psi\psi}}}=2\pi
\end{eqnarray}

For any bubble rod $[a_{2m-1},a_{2m}], m=1,2$ corresponding to an axis of the Killing vector ${\partial/\partial \phi}$
the regularity condition gives

\begin{eqnarray}
(\Delta \phi)_{Rod[a_{2m-1},a_{2m}]} = 2\pi \lim_{\rho\to 0}\sqrt{{\rho^2 g_{\rho\rho}\over g_{\phi\phi}}}= \left( {{\cal Y}\over W}\right)^{3/2}_{Rod[a_{2m-1},a_{2m}]}
(\Delta \phi)^{E}_{Rod[a_{2m-1},a_{2m}]}
\end{eqnarray}
where

\begin{eqnarray}\label{deltaphi}
(\Delta \phi)^{E}_{Rod[a_1,a_2]}&=& 4\pi (a_4- a_1)\sqrt{\frac{a_2 - a_1}{a_3 - a_1}}, \nonumber \\
(\Delta \phi)^{E}_{Rod[a_3,a_4]}&=& 4\pi (a_4- a_1)\sqrt{\frac{a_4 - a_3}{a_4 - a_2}},
\end{eqnarray}
are the periods for the seed solution corresponding to the two bubble rods, and

\begin{eqnarray}
\left({{\cal Y}\over W}\right)_{Rod[a_1,a_2]}&=&\left[ 1 + \alpha\beta\left({k_2-a_1\over k_1-a_1 }\right)\over 1 + \alpha\beta \right]^2\left({k_1-a_1\over k_2-a_1 }\right),\nonumber \\
\left({{\cal Y}\over W}\right)_{Rod[a_3,a_4]}&=&\left[ 1 + \alpha\beta \left({a_4-k_1\over a_4-k_2 }\right)\over 1 + \alpha\beta \right]^2 \left({a_4-k_2\over a_4 - k_1 }\right).\nonumber
\end{eqnarray}

Then  we  obtain two balance conditions

\begin{eqnarray}
(\Delta \phi)_{Rod[a_{2m-1},a_{2m}]} = L, \;\; m=1,2.
\end{eqnarray}

The parameters can be adjusted in appropriate way so that the balance conditions be satisfied.

\subsection{Dipole charge}

The dipole charge associated with the black ring is defined as

\begin{eqnarray}
 Q ={1\over 2\pi} \int_{S^2_{\cal H}}F
\end{eqnarray}
where $S^2_{\cal H}$ is the 2-sphere of the black ring horizon.

\bigskip
When we calculate the charge of the black ring horizon  for our solution we have to consider two separate cases, i.e. when the parameters of the soliton transformation $k_1$ and $k_2$ belong to the first bubble rod $a_1 < k_i < a_2$ or to the second bubble rod $a_3 < k_i < a_4$.
We find the following expressions

\begin{eqnarray}
&& a_1 < k_i < a_2 \nonumber \\ \\
&& Q= \frac{L\sqrt{3}\beta\Delta k (a_3-a_2) (a_2 - k_2)^{-1} (a_4-k_2)^{-1}\left[ 1 + \alpha\beta\prod^4_{i=3}\left(\frac{a_i - k_1}{a_i - k_2}\right)^{(-1)^i}\right]}{2\pi\left[ 1 + \alpha\beta\left(\frac{a_4 - k_1}{a_4 - k_2}\right)\right] \left[ 1 + \alpha\beta\prod^4_{i=2}\left(\frac{a_i - k_1}{a_i - k_2}\right)^{(-1)^i}\right]},  \nonumber \\ \\
&&a_3 < k_i < a_4 \nonumber \\ \nonumber \\
&& Q=-\frac{L \sqrt{3}\alpha \Delta k (a_3-a_2) (k_1 - a_1)^{-1} (k_1 - a_3)^{-1}\left[ 1 + \alpha\beta\prod^2_{i=1}\left(\frac{a_i - k_1}{a_i - k_2}\right)^{(-1)^i}\right]}{2\pi\left[ 1 + \alpha\beta\left(\frac{k_2 - a_1}{k_1 - a_1}\right)
\right] \left[ 1 + \alpha\beta\prod^3_{i=1}\left(\frac{a_i - k_1}{a_i - k_2}\right)^{(-1)^i}\right]}.\nonumber
\end{eqnarray}

\subsection{Dipole potential and magnetic fluxes}

If we consider the dual field $H$ defined by $H = \star F$ and more precisely
\begin{eqnarray}
i_{\zeta}i_{\xi}H= i_{\zeta}i_{\xi}\star F,
\end{eqnarray}
 we can show (see the next section) that there exists  a potential ${\cal B}$ such that
\begin{eqnarray}
i_{\zeta}i_{\xi}H=d{\cal B}.
\end{eqnarray}

In our case ${\cal B}$ can be given explicitly  by the expression

\begin{eqnarray}
{\cal B}=  \sqrt{3} e^{-u_{0}} {{\hat \omega}_{k_1k_2} \over W_1} + C_B,
\end{eqnarray}
where $C_{B}$ is a constant. The constant $C_{B}$ plays no essential role and we set it zero.
The function $\omega_{k_1k_2}$ is given by \cite{Yazadjiev}

\begin{eqnarray}
\omega_{k_1k_2} = [(R_{k_1} + R_{k_2})^2 - (\Delta k)^2](1+ab)[(R_{k_1} - R_{k_2} + \Delta k)b + (R_{k_1} - R_{k_2} -\Delta k)a] \nonumber \\
+  [(R_{k_1} - R_{k_2})^2 - (\Delta k)^2](b-a)[(R_{k_1} + R_{k_2} + \Delta k) - (R_{k_1} + R_{k_2} - \Delta k)ab].
\end{eqnarray}

The asymptotic behaviour of the potential ${\cal B}$ is

\begin{eqnarray}
{\cal B}\approx {\sqrt{3}\Delta k\over 1 + \alpha\beta }\left[(1-\cos\theta)\beta - (1+\cos\theta)\alpha\right] .
\end{eqnarray}

Further, we need the value  of the potential  ${\cal B}$ on the black ring horizon and on the axes of the Killing field $\zeta=\partial/\partial\psi$.
After some algebra we find

\begin{eqnarray}
{\cal B}_{{\cal H}}&=&-\frac{2\sqrt{3}\alpha\Delta k \prod^4_{i=3}(a_i - k_1)^{(-1)^{i}}}{
\left[ 1 + \alpha\beta\prod^4_{i=3}\left(\frac{a_i - k_1}{a_i - k_2}\right)^{(-1)^i} \right]},~~~~~ a_1 < k_i < a_2 \nonumber \\ \\
{\cal B}_{{\cal H}}&=&\frac{2\sqrt{3}\beta\Delta k \prod^2_{i=1}(k_2 - a_i)^{(-1)^{i+1}}} {\left[ 1 + \alpha\beta\prod^2_{i=1}\left(\frac{a_i - k_1}{a_i - k_2}\right)^{(-1)^i} \right]}, ~~~~~~
 a_3 < k_i < a_4  \nonumber, \\ \nonumber \\
 {\cal B}^{+}&=&{\cal B}_{Rod[z_4,+\infty]}=- 2\sqrt{3}\Delta k{\alpha\over 1 + \alpha\beta }, \\ \nonumber \\
 {\cal B}^{-}&=&{\cal B}_{Rod[-\infty,z_1]}= 2\sqrt{3}\Delta k{\beta\over 1 + \alpha\beta }.
\end{eqnarray}

The magnetic fluxes $\Psi^{+}$ and $\Psi^{-}$ are defined in the next section - see the discussion around eqs. (\ref{magnetic_flux1})
and (\ref{magnetic_flux2}). For our exact solution we find

\begin{eqnarray}
\Psi^{+}= L {\sqrt{3} \Delta k \beta (a_4-k_2)^{-1}\over \left[1 + \alpha\beta {a_4-k_1\over a_4-k_2} \right]  }, \\
\nonumber \\
\Psi^{-}=- L  {\sqrt{3} \Delta k \alpha (k_1-a_1)^{-1}\over \left[1 + \alpha\beta {k_2-a_1\over k_1-a_1} \right]  }.
\end{eqnarray}

The quantities ${\cal B}_{{\cal H}}$, ${\cal B}^{+}$, ${\cal B}^{-}$, $\Psi^{+}$ and $\Psi^{-}$  play important role in the Smarr-like relations and the mass and tension
first laws as we will see in the next section.

\subsection{Mass and tension}
The ADM mass and the tension can be calculated from the asymptotic expansion of the metric

\begin{eqnarray}
&&M=\frac{1}{4}L (2c^E_t-c^E_\phi)={L\over L^{E}}M^{E}, \\
&& {\cal T} L =\frac{1}{4}L ( c^E_t-2c^E_\phi) + \frac{3}{4}L~ \frac{1-\alpha\beta}{1+\alpha\beta}\Delta k = {\cal T}^E\frac{L}{L^E} + \frac{3}{4}L~ \frac{1-\alpha\beta}{1+\alpha\beta}\Delta k,
\end{eqnarray}
here ${\cal T}^E$ and $L^E$ are the tension and the length of the Kaluza-Klein circle at infinity corresponding to the seed solution.


\subsection{Temperature and Entropy}

The temperature of the event horizon is given by

\begin{equation}
T = \frac{1}{2\pi} \lim_{\rho\rightarrow 0}
\sqrt{ \frac{-g_{tt}}{\rho^2 g_{\rho\rho}}} \, .
\end{equation}
Applying this formula to our solution we find

\begin{equation}
T = {\cal Y}^{-3/2}_{{\cal H}}T^E,
\end{equation}
where

\begin{equation}
{\cal Y}_{{\cal H}} = \left[\frac{1+\alpha\beta\prod^4_{i=3}\left( \frac{a_i -k_1}{a_i - k_2} \right)^{(-1)^i}}{1 + \alpha\beta}\right]^2 \prod^4_{i=3}\left( \frac{a_i -k_2}{a_i - k_1} \right)^{(-1)^i} \nonumber
\end{equation}
and
\begin{eqnarray}
T^E = \frac{1}{4\pi}\frac{\sqrt{a_4 - a_2}\sqrt{a_3-a_1}}{(a_4-a_1)(a_3-a_2)}\
\end{eqnarray}
is the temperature of the event horizon for the seed solution and the metric function ${\cal Y}$ is evaluated on the horizon rod $a_2<z<a_3$.

\paragraph{}Further we can find the black ring entropy

\begin{equation}
S =  {\cal Y}^{3/2}_{{\cal H}}S^E \left(\frac{L}{L^E}\right).
\end{equation}
Again

\begin{equation}
S^E = \frac{L^E}{4T^E}(a_3-a_2)
\end{equation}
is the entropy corresponding to the seed solution.

\subsection{Mirror solutions}
The solutions with $k_i$ placed on different bubble rods are related by a discrete symmetry described below.
For a given solution with parameters $a_1, a_2, a_3, a_4, k_1, k_2$ where  $a_1<k_i<a_2$, we can find a "mirror" solution which has the same mass and opposite charge. The parameters  $a^{\,\prime}_1, a^{\,\prime}_2, a^{\,\prime}_3, a^{\,\prime}_4, k^{\,\prime}_1, k^{\,\prime}_2$ of this "mirror"  solution are given by the  transformations

\begin{eqnarray}
&&(k_2-a_1) \to (a^{\,\prime}_4 -k^{\,\prime}_1),\nonumber \\
&&(a_2 - k_1) \to (k^{\,\prime}_2 - a^{\,\prime}_3), \nonumber  \\
&& (a_4 - a_3) \to (a^{\,\prime}_2 - a^{\,\prime}_1), \\
&& (a_3 -a_2) \to (a^{\,\prime}_3 - a^{\,\prime}_2), \nonumber \\
&& (k_1 - k_2) \to (k^{\,\prime}_1 - k^{\,\prime}_2),\nonumber
\end{eqnarray}
where $a^{\,\prime}_3<k^{\,\prime}_i<a^{\,\prime}_4$. The physical quantities of the mirror solution are given by

\begin{eqnarray}
&&M^{\,\prime }=M, \,\,  L^{\,\prime }=L, \,\, {\cal T}^{\,\prime }={\cal T},  \,\, S^{\, \prime}=S, \,\, T^{\,\prime}=T, \\
&&Q^{\,\prime }=- Q, \,\, \Psi^{\,\prime\,+}=- \Psi^{-}, \,\, \Psi^{\,\prime\,-}=- \Psi^{+}, {\cal B}^{\,\prime\, +}=-{\cal B}^{-}, \,\,
{\cal B}^{\,\prime\, -}=-{\cal B}^{+}.\nonumber
\end{eqnarray}

We can also consider a mirror solution of second kind. This solution is obtained from the mirror solution by reversing the sign of the
electromagnetic potential, i.e. $\lambda^{\,\prime\prime}=-\lambda^{\,\prime}=-\lambda$. For the mirror solution of the second kind
we have

\begin{eqnarray}
&&M^{\,\prime \prime}=M, \,\,  L^{\,\prime \prime }=L, \,\, {\cal T}^{\,\prime\prime }={\cal T},  \,\, S^{\, \prime\prime}=S, \,\, T^{\,\prime\prime}=T, \\
&&Q^{\,\prime \prime}= Q, \,\, \Psi^{\,\prime\prime\,+}= \Psi^{-}, \,\, \Psi^{\,\prime\prime\,-}= \Psi^{+}, {\cal B}^{\,\prime\prime\, +}={\cal B}^{-}, \,\,{\cal B}^{\,\prime\prime\, -}={\cal B}^{+}.\nonumber
\end{eqnarray}

\subsection{Parameter counting}
Our solutions are characterized by 4 parameters - the lengths of the three finite rods and $\Delta k$. For a given length $L$ of the KK circle at infinity we have 2 constraints coming from the balance conditions. Therefore we are left with 2 parameters for the regular solutions.
This means that the regular solutions are characterized by two independent parameters
which can be chosen to be the mass $M$ and the dipole charge $Q$. Let us note, however, that the solutions in the general case are not uniquely
specified by the mass and the dipole charge. There are different solutions which can have the same mass and dipole charge -- for example, a given
solution and its mirror solution of second kind have the same mass and dipole charge. 

The full classification of the KK black holes with dipole charges exceeds the scope of this paper. This question was briefly discussed
in \cite{YazadjievNedkova} where it was pointed out that the dipole KK black holes (with electromagnetic field along the compact dimension) can classified by the rod structure, the dipole charges and  also by the  magnetic flux(es). Detailed consideration of the classification will be presented elsewhere.

\section{Smarr-like relations and first laws for the mass and the tension }
In this section we derive the Smarr-like relations and first laws for the mass and the tension for the configurations under consideration.
Our derivation will be done for the more general case of 5D Einstein-Maxwell-dilaton gravity  given by the field equations

\begin{eqnarray}\label{EMDFE}
&&R_{\mu\nu}= 2\partial_{\mu}\varphi\partial_{\nu}\varphi +{1\over 2}e^{-2\gamma\varphi}\left(F_{\mu\sigma}F_{\nu}^{\,\sigma}
- {1\over 6}g_{\mu\nu}F_{\lambda\sigma}F^{\lambda\sigma}\right), \nonumber \\ \nonumber \\
&&\nabla_{\mu}\left(e^{-2\gamma\varphi}F^{\mu\nu} \right)=0 , \;\; \nabla_{[\sigma} F_{\mu\nu]}=0, \\ \nonumber \\
&&\nabla_{\mu}\nabla^{\mu}\varphi = - {\gamma\over 8}e^{-2\gamma\varphi}F_{\sigma\lambda}F^{\sigma\lambda}, \nonumber
\end{eqnarray}
where $R_{\mu\nu}$ is the Ricci tensor for the spacetime metric $g_{\mu\nu}$, $F_{\mu\nu}$ is
the Maxwell tensor, $\varphi$ is the dilaton field and $\gamma$ is the dilaton coupling parameter. For $\gamma=0$ (and $\varphi=0$)
we obtain the 5D Einstein-Maxwell equations.

\subsection{Smarr-like relations}

In order to derive the Smarr-like relations we shall use the generalized Komar integrals \cite{Townsend:2001rg}
in the form presented in \cite{YazadjievNedkova}.  The generalized Komar integrals are given by

\begin{eqnarray}
M =  - {L\over 16\pi} \int_{S^{2}_{\infty}} \left[2i_\eta \star d\xi - i_\xi \star d\eta \right],
\end{eqnarray}

\begin{eqnarray}
{\cal T}=  - {1\over 16\pi} \int_{S^{2}_{\infty}} \left[i_\eta \star d\xi - 2i_\xi \star d\eta \right],
\end{eqnarray}
where the integration is performed over the 2-dimensional sphere  at the spatial infinity of ${\cal M}^{4}$.

The generalized Komar integrals allow us to define the intrinsic mass of  each object in the configuration \cite{KY}.
The intrinsic mass of each black hole  is given by

\begin{eqnarray}\label{BHKOMMARMASS}
M^{{\cal H}}_{i}= - {L\over 16\pi} \int_{{\cal H}_{i}} \left[2i_\eta \star d\xi - i_\xi \star d\eta \right]
\end{eqnarray}
where ${\cal H}_i$ is the 2-dimensional surface which is an intersection of the i-th horizon with a constant  $t$ and $\phi$
hypersurface. Analogously the intrinsic mass of each bubble is

\begin{eqnarray}\label{BKOMMARMASS}
M^{{\cal B}}_{j}= - {L\over 16\pi} \int_{{\cal B}_{j}} \left[2i_\eta \star d\xi - i_\xi \star d\eta \right].
\end{eqnarray}

One can show that the intrinsic masses of the black holes and bubbles are given by
\begin{eqnarray}
M^{{\cal H}}_{i}= {1\over 2}L l^{{\cal H}}_i, \; \; \;
M^{{\cal B}}_{j}= {1\over 4} L l^{{\cal B}}_j,
\end{eqnarray}
where $l^{{\cal H}}_i$ and $l^{{\cal B}}_j$ are the lengths of the horizon and bubble rods, respectively. It was also  shown in \cite{KY} that

\begin{eqnarray}
M^{{\cal H}}_{i}= {1\over 4\pi} \kappa_{{\cal H}_i} {\cal A}_{{\cal H}_i}, \;\;\;
M^{\cal B}_{j}= {L\over 8\pi}\kappa_{{\cal B}_j} {\cal A}_{{\cal B}_j},
\end{eqnarray}
where $\kappa_{{\cal H}_i}$ and ${\cal A}_{{\cal H}_i}$ are the surface gravity and the area of the i-th horizon and the surface gravity and area of j-th bubble. The surface gravity and the area for a bubble were first introduced  in \cite{Kastor:2008wd}. The bubble surface gravity is defined by

\begin{eqnarray}
\kappa^2_{\cal B}= {1\over 2} \nabla_{[\mu}\eta_{\nu]} \nabla^{[\mu}\eta^{\nu]}
\end{eqnarray}
where the right hand side is evaluated on the bubble. The reader might consult \cite{Kastor:2008wd} for other equivalent definitions. The bubble area is given by

\begin{eqnarray}
{\cal A}_{\cal B}= \int_{{\cal B}}\sqrt{|g_{tt}| g_{\rho\rho}g_{\psi\psi}}\,dzd\psi.
\end{eqnarray}

For regular (smooth) bubbles (i.e. bubbles without conical singularities), the case we consider here, one can show that

\begin{eqnarray}
\kappa_{\cal B}= {2\pi\over  L}.
\end{eqnarray}

Using  Stokes theorem the tension can be represented as a bulk integral over a constant t and $\phi$ hypersurface $\Sigma$ and surface integrals over the black hole horizons and bubbles

\begin{eqnarray}
{\cal T} L &=& -{L\over 16\pi} \sum_i\int_{{\cal H}_i} \left(i_\eta \star d\xi - 2i_\xi\star d\eta\right)
-{L\over 16\pi} \sum_j\int_{{\cal B}_j} \left(i_\eta \star d\xi - 2i_\xi\star d\eta\right) \\
&&-{L\over 16\pi} \int_{\Sigma}d\left(i_\eta \star d\xi - 2i_\xi\star d\eta \right) \nonumber
\end{eqnarray}
where we have taken into account that $\partial\Sigma =S^{2}_{\infty} -\sum_i {\cal H}_i - \sum_j {\cal B}_j$. Using the definitions
(\ref{BHKOMMARMASS}) and (\ref{BKOMMARMASS}), the Killing symmetries and the identity $d\star d\xi=2\star R[\xi]$ for an arbitrary
Killing field, we have

\begin{eqnarray}
{\cal T} L = {1\over 2} \sum_i M^{{\cal H}}_i + 2\sum_j M^{{\cal B}}_j + {L\over 8\pi} \int_{\Sigma} \left(i_{\eta}\star R[\xi] - 2i_\xi \star R[\eta] \right)
\end{eqnarray}
where $R[X]$ is the Ricci 1-form\footnote{We recall that the Ricci 1-form $R[X]$ is defined by $R[X]=R_{\mu\nu} X^{\mu}dx^\nu.$} with respect to the vector field $X$. Making advantage of the field equations (\ref{EMDFE}) we obtain

\begin{eqnarray}
\star R[\xi] = - {1\over 2}e^{-2\gamma\varphi} \left( -{2\over 3}i_{\xi}F\wedge \star F + {1\over 3} F\wedge i_{\xi}\star F \right)
\end{eqnarray}
and the same expression for $\star R[\eta]$, however with $\xi$ replaced by $\eta$. Hence we find

\begin{eqnarray}
i_{\eta}\star R[\xi] - 2i_\xi \star R[\eta]= {1\over 2}e^{-2\gamma\varphi} i_\eta F \wedge i_\xi\star F
\end{eqnarray}
and therefore

\begin{eqnarray}\label{Smarr0}
{\cal T} L = {1\over 2} \sum_i M^{{\cal H}}_i + 2\sum_j M^{{\cal B}}_j + {L\over 16\pi} \int_{\Sigma} e^{-2\gamma\varphi} i_\eta F \wedge i_\xi\star F.
\end{eqnarray}

Using now the invariance under the Killing field $\zeta$ we can write

\begin{eqnarray}
\int_{\Sigma} e^{-2\gamma\varphi} i_\eta F \wedge i_\xi\star F=2\pi \int_{\hat {\cal M}} i_{\zeta}\left[e^{-2\gamma\varphi} i_\eta F \wedge i_\xi\star F \right]
\end{eqnarray}
where $\hat {\cal M}={\cal M}/U(1)^2\times R=\Sigma/U(1)$ is the factor space. For $i_{\zeta}i_{\eta} F$ we have

\begin{eqnarray}
di_{\zeta}i_{\eta} F=i_{\zeta}i_{\eta}dF +i_\xi{\cal L}_{\eta}F - i_\eta{\cal L}_{\xi}F=0
\end{eqnarray}
and taking into account that $i_{\zeta}i_{\eta} F$ vanishes on the axes of $\eta$ and $\zeta$ we conclude that $i_{\zeta}i_{\eta} F=0$ everywhere.
Hence we find

 \begin{eqnarray}
\int_{\Sigma} e^{-2\gamma\varphi} i_\eta F \wedge i_\xi\star F=
2\pi \int_{\hat {\cal M}} i_{\zeta}\left[e^{-2\gamma\varphi} i_\eta F \wedge i_\xi\star F \right]= \\ -2\pi \int_{\hat {\cal M}}
e^{-2\gamma\varphi} i_\eta F \wedge i_{\zeta}i_\xi\star F. \nonumber
\end{eqnarray}

As a consequence of the field equations and the spacetime symmetries we have $d[e^{-2\alpha\varphi}i_{\zeta}i_\xi\star F]=0$.
Since the factor space is simply connected \cite{HY} there exists a globally well-defined potential ${\cal B}$ on $\hat {\cal M}$
such that $e^{-2\alpha\varphi}i_{\zeta}i_\xi\star F=d{\cal B}$. With this in mind we obtain

\begin{eqnarray}
\int_{\Sigma} e^{-2\gamma\varphi} i_\eta F \wedge i_\xi\star F=-2\pi \int_{\hat {\cal M}}
e^{-2\gamma\varphi} i_\eta F \wedge i_{\zeta}i_\xi\star F=  \\ 2\pi \int_{\hat {\cal M}} d[{\cal B}i_{\eta}F ]=
2\pi \int_{\partial \hat {\cal M}} {\cal B} i_{\eta} F. \nonumber
\end{eqnarray}

The next step is to calculate the integral on the boundary of the factor space which formally can be presented in the following way
$\partial \hat {\cal M}= Arc(\infty) + Rod_{\psi}(-\infty, z_1]  + \sum_i Rod-Horizon_i  + \sum_j Rod-Bubble_j + Rod_{\psi}[z_{2N},+\infty)$,
where $Arc(\infty)$ is the upper infinite semi-circle. It can be shown that for an upper semi-circle with radius $R$ we have

\begin{eqnarray}
\int_{Arc(R)} {\cal B} i_{\eta} F \sim {1\over R}
\end{eqnarray}
which means that this integral does not give any contribution in the limit $R\to \infty$.
The same is true for the integral on the bubble rods since the Killing field $\eta$ vanishes on them.
In this way we find

\begin{eqnarray}
\int_{\partial \hat {\cal M}} {\cal B} i_{\eta} F = \int_{ Rod_{\psi}(-\infty, z_1]} {\cal B} i_{\eta} F +
\int_{ Rod_{\psi}[z_{2N},\infty]} {\cal B} i_{\eta} F + \sum_i \int_{ Rod-Horizon_i} {\cal B} i_{\eta} F.
\end{eqnarray}

One can prove that the potential ${\cal B}$ is constant on the horizon rods, therefore

\begin{eqnarray}
\sum_i \int_{ Rod-Horizon_i} {\cal B} i_{\eta} F= \sum_i {\cal B}_i\int_{ Rod-Horizon_i}  i_{\eta} F=
 \sum_i {{\cal B}_i\over L} \int_{S^2_{H_i}} F= \sum_i {2\pi {\cal B}_i\over L} Q_{i}
\end{eqnarray}
where

\begin{eqnarray}
Q_i= {1\over 2\pi} \int_{S^2_{H_i}} F
\end{eqnarray}
is the magnetic (dipole) charge associated with the i-th horizon. In order to prove that ${\cal B}$ is constant on the horizons
we consider the following chain of equalities

\begin{eqnarray}
<e^{-2\gamma\varphi} i_{\zeta}i_{\xi}\star F,e^{-2\gamma\varphi} i_{\zeta}i_{\xi}\star F> =
-e^{-4\gamma\varphi} <\xi \wedge (F\wedge \zeta),\xi \wedge (F\wedge \zeta)> = \nonumber \\
- e^{-4\gamma\varphi}<\xi,\xi> <F\wedge \zeta, F\wedge \zeta> + e^{-4\gamma\varphi}<i_{\xi}(F\wedge \zeta),i_{\xi}(F\wedge \zeta)>
\end{eqnarray}
where $<,>$ denotes the inner product of two forms of the same degree. Further we should take into account that $i_{\xi}(F\wedge \zeta)=0$
since Killing fields $\xi$ and $\zeta$ are orthogonal and in our case $i_{\xi}F=0$.  In this way we find

\begin{eqnarray}
<e^{-2\gamma\varphi} i_{\zeta}i_{\xi}\star F,e^{-2\gamma\varphi} i_{\zeta}i_{\xi}\star F>=- e^{-4\gamma\varphi}<\xi,\xi> <F\wedge \zeta, F\wedge \zeta>=\\ - e^{-4\gamma\varphi}<\xi,\xi><\zeta,\zeta><F,F> \nonumber
\end{eqnarray}
which shows that $e^{-2\gamma\varphi} i_{\zeta}i_{\xi}\star F$ is null on the horizons where $<\xi,\xi>=g(\xi,\xi)=0$. Taking into account also that
 $e^{-2\gamma\varphi} i_{\zeta}i_{\xi}\star F$ is orthogonal to $\xi$ (by definition) we conclude that
$e^{-2\gamma\varphi} i_{\zeta}i_{\xi}\star F$ is proportional to $\xi$ on the horizons, i.e. $e^{-2\gamma\varphi} i_{\zeta}i_{\xi}\star F=\Lambda \xi$ (on the horizons). For arbitrary vector field $u$ tangent to a given horizon we have $i_u d{\cal B}={\cal L}_{u}{\cal B}=\Lambda i_u\xi=0 $ which
shows that ${\cal B}$ is indeed constant on the horizons.

Let us now consider the integrals on the semi-infinite rods of the axis of $\zeta$. The potential ${\cal B}$ is
constant on the axes of $\zeta$ and this follows directly from the definition of the potential ${\cal B}$.  To be specific we will consider $Rod_{\psi}[a_{2N},+\infty)$.
In the next step we follow \cite{YazadjievNedkova} and define $C^{+}$ to be the 2-dimensional surface generated from the path $[a_{2N},\infty)$
by acting with the isometry generated by $\eta$. Since $\eta|_{a_{2N}}=0$ the 2-surface $C^{+}$ has disk topology. Then we have

\begin{eqnarray}
\int_{ Rod_{\psi}[z_{2N},\infty]} {\cal B} i_{\eta} F= {\cal B}^{+} \int_{ Rod_{\psi}[a_{2N},\infty)} i_{\eta} F=
{{\cal B}^{+} \over L} \int_{C^{+}} F=  {{\cal B}^{+} \over L} \Psi^{+}
\end{eqnarray}
where ${\cal B}^{+}={\cal B}_{|_{Rod_{\psi}[a_{2N},+\infty) }}$ and we have introduced the magnetic flux through the 2-surface $C^{+}$ defined by

\begin{eqnarray}\label{magnetic_flux1}
\Psi^{+}= \int_{C^{+}} F.
\end{eqnarray}

Analogously  we obtain
\begin{eqnarray}
\int_{ Rod_{\psi}(-\infty, a_1]} {\cal B} i_{\eta} F=  {{\cal B}^{-} \over L} \Psi^{-}
\end{eqnarray}
where

\begin{eqnarray}\label{magnetic_flux2}
\Psi^{-}= \int_{C^{-}} F
\end{eqnarray}
is the magnetic flux through the 2-dimensional surface $C^{-}$ generated from $(-\infty,a_1]$ by acting with the isometry generated by
$\eta$ and ${\cal B}^{-}={\cal B}_{|_{Rod_{\psi}[-\infty,a_1)}}$.

Summarizing the results so far we obtain

\begin{eqnarray}
\int_{\Sigma} e^{-2\alpha\varphi} i_\eta F \wedge i_\xi\star F=
2\pi \int_{\partial \hat {\cal M}} {\cal B} i_{\eta} F= \sum_i 4\pi^2 {\cal B}_i {Q_i\over L} + 2\pi {\cal B}^{+}{\Psi^{+}\over L}
+2\pi {\cal B}^{-}{\Psi^{-} \over L}
\end{eqnarray}
or equivalently

\begin{eqnarray}\label{TSmarr0}
{\cal T}L= {1\over 2} \sum_i M^{{\cal H}}_i + 2 \sum_j M^{{\cal B}}_j + \sum_i {\pi\over 4} {\cal B}_{i} Q_i +
{1\over 8} {\cal B}^{+} \Psi^{+} + {1\over 8} {\cal B}^{-} \Psi^{-} .
\end{eqnarray}

The magnetic fluxes $\Psi^{+}$ and $\Psi^{-}$ are related via the equation

\begin{eqnarray}\label{constraint_charge}
\Psi^{+} + \Psi^{-}= - 2\pi \sum_i Q_i.
\end{eqnarray}
This follows from the following chain of equalities

\begin{eqnarray}
0=L\int_{{\hat {\cal M}}} di_{\eta}F= L\int_{\partial{\hat {\cal M}}}i_{\eta}F=
L\int_{Rod(-\infty,a_1]}i_{\eta}F + L\sum_i \int_{Rod_{Horizon_i}}i_{\eta}F     \\ + L\sum_j \int_{Rod_{Bubble_j}}i_{\eta}F +
L\int_{Rod[a_{2N},+\infty)}i_{\eta}F= \Psi^{+} + 2\pi \sum_i Q_i + \Psi^{-} . \nonumber
\end{eqnarray}

Using this constraint and defining

\begin{eqnarray}\label{flux_def}
\Psi= {1\over 2}(\Psi^{+} - \Psi^{-})
\end{eqnarray}
we find

\begin{eqnarray}
{L\over 16\pi}\int_{\Sigma} e^{-2\gamma\varphi} i_\eta F \wedge i_\xi\star F=
2\pi \int_{\partial \hat {\cal M}} {\cal B} i_{\eta}F = \\ \sum_i {\pi\over 4} ({\cal B}_i -{1\over 2}{\cal B}^{+} - {1\over 2}{\cal B}^{-} ) Q_i +
{1\over 8}({\cal B}^{+} - {\cal B}^{-})\Psi \nonumber
\end{eqnarray}

which substituted in (\ref{Smarr0}) gives the desired Smarr-Like relation for the tension, namely

\begin{eqnarray}
{\cal T} L = {1\over 2} \sum_i M^{H}_{i} + 2\sum_j  M^{B}_{j} + \sum_i {\pi\over 4} ({\cal B}_i -{1\over 2}{\cal B}^{+} - {1\over 2}{\cal B}^{-} ) Q_i +
{1\over 8}({\cal B}^{+} - {\cal B}^{-})\Psi.
\end{eqnarray}

Following the same method as for the tension we find

\begin{eqnarray} M= \sum_i M^{{\cal H}} + \sum_j M^{{\cal B}}_j - {L\over 16\pi} \int_{\Sigma} e^{-2\alpha\varphi}i_\xi F \wedge i_\eta \star F .
\end{eqnarray}
which in view of the fact that $i_\xi F=0$ gives the Smarr-like relation for the mass

\begin{eqnarray}\label{Smarr-MASS}
M= \sum_i M^{{\cal H}}_{i} + \sum_j M^{{\cal B}}_j.
\end{eqnarray}

The Smarr-like relations for the mass and the tension were derived for the rod structure shown on fig.(\ref{rodstrn1}).
It is not difficult one to show that the derived relations also hold for more general rod structures containing black rings and bubbles.

We have checked explicitly that the derived Smarr-like relations are satisfied for the exact solution constructed in
the present paper.

\subsection{Mass and tension first law}

Our next goal is to derive the mass and tension first laws for the black  configurations under consideration.
In our derivation we shall follow in part our previous work \cite{YazadjievNedkova} based on Wald's approach \cite{Wald}.

Our diffeomorphism covariant theory is derived from  the  Lagrangian
\begin{eqnarray}
 {\mathbf  L} = \star R - 2d\varphi \wedge \star d\phi - {1\over 2}e^{-2\gamma\varphi} F\wedge \star F.
\end{eqnarray}
When the field equations are satisfied, the first order variation of the Lagrangian is given by

\begin{eqnarray}
\delta {\mathbf L} = d {\Theta}
\end{eqnarray}
where

\begin{eqnarray}
d\Theta = d\star \,\upsilon - 4 (d\star d\varphi)\delta \varphi - \left(e^{-2\gamma\varphi} \star F\right) \wedge \delta F
\end{eqnarray}
and

\begin{eqnarray}
\upsilon_\mu = \nabla^{\nu}\delta g_{\mu\nu} - g^{\alpha\beta}\nabla_{\mu}\delta g_{\alpha\beta}.
\end{eqnarray}
Here  $\delta$ denotes the first order variation of the corresponding quantity.

The Noether current ${\cal I}^{X}$ associated  with a diffeomorphism generated by an arbitrary smooth vector field $X$,
as it has been shown in \cite{Wald},  is

\begin{eqnarray}
{\cal I}^{X}= \Theta({\cal P}, {\cal L}_{X}\Gamma) - i_X {\mathbf L},
\end{eqnarray}
where the fields $g_{\mu\nu}, F, \varphi$ are collectively denoted by ${\cal P}$ . The current ${\cal I}^{X}$ satisfies
$d{\cal I}^{X}=0$ when the field equations are satisfied. Since ${\cal I}^{X}$ is closed there exists a 3-form ${\cal N}^{X}$ (Noether charge 3-form)
such that ${\cal I}=d{\cal N}^{X}$.

Now, let ${\cal P}$ be a solution to the field equations (\ref{EMDFE}) and let $\delta {\cal P}$ be a
linearized perturbation satisfying  the linearized equations of the Einstein-Maxwell-dilaton gravity.
For simplicity we will also assume that ${\cal L}_{\xi}\delta {\cal P}={\cal L}_{\eta}\delta {\cal P}={\cal L}_{\zeta}\delta {\cal P}=0$.  Then, choosing $X$ to be a Killing field one can show that \cite{Wald}

\begin{eqnarray}
\delta d{\cal N}^{X}=di_X\Theta.
\end{eqnarray}

In the case under consideration we need the Noether forms ${\cal N}^{\xi}$ and ${\cal N}^{\eta}$. After some calculations it can be shown that
they are given by

\begin{eqnarray}
&&{\cal N}^{\xi}= - \star d\xi, \\  \nonumber \\
&&{\cal N}^{\eta}= - \star d\eta - \lambda \left(e^{-2\gamma\varphi}\star F \right),
\end{eqnarray}
where the potential $\lambda$ is defined by $i_{\eta}F=-d\lambda$ (see eq. (\ref{potential_lambda})).

In fact what we need are the 2-forms $i_\eta {\cal N}^{\xi}$  and $i_\xi {\cal N}^{\eta}$ \cite{YazadjievNedkova}. For them one can show that

\begin{eqnarray}\label{Noetheridentities}
\delta \left( di_{\eta}{\cal N}^{\xi}\right)=di_{\eta}i_{\xi}\Theta, \;\;\; \; \; \delta \left(di_{\xi}{\cal N}^{\eta}\right)= - di_{\eta}i_{\xi}\Theta.
\end{eqnarray}

It turns out useful to combine (\ref{Noetheridentities}) to a single equality

\begin{eqnarray}\label{combinedidentity}
 \delta  \left(2 di_{\eta}{\cal N}^{\xi} - di_{\xi}{\cal N}^{\eta} \right)= 3di_{\eta}i_{\xi}\Theta . \end{eqnarray}

Integrating on $\Sigma$ we have

\begin{eqnarray}\label{combinedidentity1}
\delta \int_{\Sigma} \left(2 di_{\eta}{\cal N}^{\xi} - di_{\xi}{\cal N}^{\eta} \right)=
3\int_{ \Sigma} di_{\eta}i_{\xi}\Theta.
\end{eqnarray}

Calculations very similar to those in deriving the Smarr-like relations give the following result for the integral on the  left hand side of
(\ref{combinedidentity1})

\begin{eqnarray}
\int_{\Sigma} \left(2 di_{\eta}{\cal N}^{\xi} - di_{\xi}{\cal N}^{\eta} \right)=
-4\pi^2 \sum_i {\cal B}_i{Q_i\over  L} - 2\pi {\cal B}^{+}{\Psi^{+} \over L}   - 2\pi {\cal B}^{-}{\Psi^{-} \over L}
\end{eqnarray}

Respectively, for the integral of the right hand side of  (\ref{combinedidentity1}), we obtain  (see also \cite{YazadjievNedkova})

\begin{eqnarray}
\int_{ \Sigma} di_{\eta}i_{\xi}\Theta= - 4\pi (\delta c_t - \delta c_\phi) + 16\pi {\cal D}\delta\varphi_{\infty}
- 2\sum_i {{\cal A}_{H_i}\over L}\delta\kappa_{H_i} - 2\sum_j {\cal A}_{B_j} \delta \kappa_{B_j} \nonumber \\
- 4\pi^2 \sum_i {\cal B}_i \delta \left({Q_i\over L }\right) -2\pi {\cal B}^{+} \delta \left({\Psi^{+}\over L }\right)
 -2\pi {\cal B}^{-} \delta \left({\Psi^{-}\over L }\right).
\end{eqnarray}
where ${\cal D}$ is the dilaton charge defined by
\begin{eqnarray}
{\cal D}= {1\over 4\pi} \int_{S^2_{\infty}} i_\eta i_\xi\star d\varphi.
\end{eqnarray}
It is worth noting that the dilaton charge is not an independent characteristic. One can show that the dilaton
charge can be  expressed  in the form

\begin{eqnarray}
{\cal D} = -{\gamma\over 4\pi L} \left[ 2\pi\sum_i {\cal B}_i {\cal Q}_i + {\cal B}^{+}\Psi^{+} + {\cal B}^{-}\Psi^{-}\right]=\nonumber \\
-{\gamma\over 4\pi L}\left[ 2\pi\sum_i \left({\cal B}_i - {1\over 2}{\cal B}^{+} - {1\over 2}{\cal B}^{-}\right) {\cal Q}_i +
\left({\cal B}^{+} -{\cal B}^{-}\right)\Psi \right].
\end{eqnarray}

Substituting  these results in (\ref{combinedidentity1}) we obtain

\begin{eqnarray}
&& {3\over 4} (\delta c_t - \delta c_\phi)= 3{\cal D}\delta\varphi_{\infty}
- {3\over 8\pi} \sum_i {{\cal A}_{{\cal H}_i}\over L}\delta \kappa_{{\cal H}_i}
- {3\over 8\pi} \sum_j {{\cal A}_{{\cal B}_j}\over L}\delta \kappa_{{\cal B}_j} \nonumber \\
&&- {\pi\over 2} \sum_i {\cal B}_i \delta \left({Q_i\over L}\right) + {\pi\over 4} \sum_i {Q_i\over L} \delta {\cal B}_i
- {1\over 4}{\cal B}^{+}\delta \left({\Psi^{+}\over L}\right) + {1\over 8}  {\Psi^{+}\over L}\delta {\cal B}^{+}  \\
&&- {1\over 4}{\cal B}^{-}\delta \left({\Psi^{-}\over L}\right) + {1\over 8}  {\Psi^{-}\over L}\delta {\cal B}^{-} .\nonumber
\end{eqnarray}

The next step is to take into account that $3/4(\delta c_t - \delta c_\phi)=\delta (M/L) + \delta {\cal T}$ and
to express $\delta {\cal T}$ from the Smarr-like relation (\ref{TSmarr0}) which gives

\begin{eqnarray}\label{EQUALITY1}
\delta \left({M\over L}\right) &=& 3{\cal D} \delta\varphi_{\infty} - {1\over 2\pi} \sum_{i} {{\cal A}_{{\cal H}_i}\over L } \delta \kappa_{{\cal H}_i}
- {1\over 8\pi}\sum_i \kappa_{{\cal H}_i} \delta  \left({{\cal A}_{{\cal H}_i}\over L }\right)  -
 {5\over 8\pi} \sum_{j} {\cal A}_{{\cal B}_j} \delta \kappa_{{\cal B}_j} \nonumber \\
&&- {1\over 4\pi}\sum_j \kappa_{{\cal B}_j} \delta {\cal A}_{{\cal B}_j}  - {3\over 4}\pi \sum_i {\cal B}_i \delta \left({Q_i\over L}\right)
- {3\over 8} {\cal B}^{+}\delta \left({\Psi^{+}\over L}\right)  - {3\over 8} {\cal B}^{-}\delta \left({\Psi^{+}\over L}\right). \nonumber \\
 \end{eqnarray}

Now using again the Smarr-like relation (\ref{Smarr-MASS}) we also find

\begin{eqnarray}\label{EQUALITY2}
\delta \left({M\over L}\right) &=&   {1\over 4\pi} \sum_{i} {{\cal A}_{{\cal H}_i}\over L } \delta \kappa_{{\cal H}_i}
+ {1\over 4\pi}\sum_i \kappa_{{\cal H}_i} \delta  \left({{\cal A}_{{\cal H}_i}\over L }\right)  +
 {1\over 8\pi} \sum_{j} {\cal A}_{{\cal B}_j} \delta \kappa_{{\cal B}_j} \\
&&+ {1\over 8\pi}\sum_j \kappa_{{\cal B}_j} \delta {\cal A}_{{\cal B}_j} . \nonumber
\end{eqnarray}

Combining the above equalities (\ref{EQUALITY1}) and (\ref{EQUALITY2}) we obtain

\begin{eqnarray}
\delta \left( {M\over L}  \right) = {\cal D}\delta\varphi_{\infty} + \sum_i {\kappa_{{\cal H}_i}\over 8\pi} \delta \left({{\cal A}_{{\cal H}_{i}}\over L}\right)
- {1\over 8\pi}\sum_j {\cal A}_{{\cal B}_j} \delta \kappa_{{\cal B}_j} - {\pi\over 4} \sum_i {\cal B}_i \delta \left({Q_i\over L}\right) \nonumber \\
-  {1\over 8} {\cal B}^{+}\delta \left({\Psi^{+}\over L}\right)  -  {1\over 8} {\cal B}^{-}\delta \left({\Psi^{-}\over L}\right)
\end{eqnarray}
which in view of the relation (\ref{TSmarr0}) gives

\begin{eqnarray}\label{MASSFIRSTLAW}
\delta M = L{\cal D}\delta\varphi_{\infty}  + \sum_i {\kappa_{{\cal H}_i}\over 8\pi} \delta {\cal A}_{{\cal H}_{i}}
- {1\over 8\pi} \sum_j {\cal A}_{{\cal B}_{j}}\delta (\kappa_{{\cal B}_j}L) - \sum_i {\pi\over 4 }{\cal B}_i \delta Q_i
\nonumber \\ - {1\over 8}{\cal B}^{+}\delta \Psi^{+}  - {1\over 8}{\cal B}^{-}\delta \Psi^{-} + {\cal T}\delta L.
\end{eqnarray}

Further taking into account (\ref{constraint_charge}) and (\ref{flux_def}) we find

\begin{eqnarray}\label{MASSFIRSTLAWLAST}
&&\delta M = L{\cal D}\delta\varphi_{\infty}  + \sum_i {\kappa_{{\cal H}_i}\over 8\pi} \delta {\cal A}_{{\cal H}_{i}}
- {1\over 8\pi} \sum_j {\cal A}_{{\cal B}_{j}}\delta (\kappa_{{\cal B}_j}L) - \sum_i {\pi\over 4}\left( {\cal B}_i - {1\over 2} {\cal B}^{+}
 - {1\over 2} {\cal B}^{-}\right) \delta Q_i \nonumber \\
&& -{1\over 8} \left({\cal B}^{+} - {\cal B}^{-} \right)\delta \Psi + {\cal T} \delta L .
\end{eqnarray}

This is the desired form of the mass first law. For smooth bubbles (the case we consider here) $\delta (\kappa_{{\cal B}_j}L)=0$ and the third
term in (\ref{MASSFIRSTLAWLAST}) gives no contribution.  Once having the mass first law, the tension first law can be easily found and the result is

\begin{eqnarray}  \delta {\cal T}= {\cal D}\delta \varphi_{\infty} + {1\over 8\pi} \sum_j \kappa_{{\cal B}_j}\delta {\cal A}_{{\cal B}_j}
- {1\over 8\pi}\sum_i{{\cal A}_{{\cal H}_i}\over  L} \delta \kappa_{{\cal H}_i} \nonumber \\ +  \sum_i {\pi\over 4} {Q_i\over L} \delta \left({\cal B}_i - {1\over 2}{\cal B}^{+} - {1\over 2}{\cal B}^{-}\right)
+ {\Psi\over L} \delta\left[{1\over 8}({\cal B}^{+} - {\cal B}^{-})\right].
\end{eqnarray}

The mass and the tension first laws we derived in this subsection also hold in the general case not only for the rod structure
shown in fig. (\ref{rodstrn1}).

We have checked explicitly that the mass and tension first laws are  satisfied for our exact solutions.

\section{Conclusion}

In the present paper we have constructed new exact solutions to 5D Einstein-Maxwell gravity describing sequences of Kaluza-Klein bubbles
and dipole black rings. The basic properties and characteristics of the solutions were calculated and discussed.
We  also derived  the Smarr-like relations and the mass and tension first laws. The novel feature is the appearance of the magnetic flux in the
Smarr like relations and the first laws. Another interesting feature is the fact that  the effective  potential associated with the dipole
charge involves not only the value of ${\cal B}$ on the horizons but also the values of  ${\cal B}$ on the axes  of the non-compact direction.

Future work may involve the inclusion of rotation.

\section*{Acknowledgements}
 The partial support by the Bulgarian National Science Fund under Grants  VUF-201/06, DO 02-257 and
Sofia University Research Fund under Grant No 074/2009, is gratefully  acknowledged.

\appendix

\setcounter{equation}{0}
\renewcommand{\theequation}{\Alph{section}.\arabic{equation}}

\section{Multi black rings and Kaluza-Klein bubbles \\ sequences}

\label{app:general}

The general form of the vacuum solution describing sequences of $q$ dipole black rings and $p=q+1$ Kaluza-Klein bubbles  is
\cite{EHO}

\begin{eqnarray}
&&g^{E}_{00}= -\prod^{N-1}_{i=2} (R_{a_i}-\zeta_{a_i})^{(-1)^{i}}=-\prod^{N-1}_{i=2} \left(e^{2U_{a_i}}\right)^{(-1)^{i}}, \\ \nonumber \\
&&g^{E}_{\phi\phi}= \prod_{i=1}^{N} (R_{a_i}-\zeta_{a_i})^{(-1)^{i+1}}=\prod_{i=1}^{N} \left(e^{2{\tilde U}_{a_i}}\right)^{(-1)^{i}}, \\ \nonumber \\
&&g^{E}_{\psi\psi}= (R_{a_1} + \zeta_{a_1})(R_{a_N}- \zeta_{a_N}), \\ \nonumber \\
&&g^{E}_{\rho\rho}= {Y_{1N}\over 2^{N/2} }\left(\prod_{i=1}^{N} {1\over R_i}\right)\left(\prod_{2\le i<j\le N-1} Y^{(-1)^{i+j+1}}_{ij}\right)\sqrt{\prod_{i=2}^{N-1} \left({Y_{1i}\over Y_{iN}}\right)^{(-1)^{i}} } {R_{a_N}-\zeta_{a_N}\over R_{a_1}-\zeta_{a_1}},
\nonumber \\
\end{eqnarray}
where

\begin{eqnarray}
Y_{ij}= R_{a_i}R_{a_j} +\zeta_{a_i}\zeta_{a_j} + \rho^2 .
\end{eqnarray}

\begin{eqnarray}
&&a=\alpha \prod^{N}_{i=1} \left( {e^{2U_{k_1}} + e^{2{\tilde U}_{a_i}}}\over e^{{\tilde U}_{a_i}}\right)^{(-1)^{i}} \\ \nonumber\\
&&b= \beta \prod^{N}_{i=1} \left( {e^{2U_{k_2}} + e^{2{\tilde U}_{a_i}}}\over e^{{\tilde U}_{a_i}}\right)^{(-1)^{i+1}}
\end{eqnarray}

The 2-soliton transformation applied to this seed solution produces a solution to the 5D Einstein-Maxwell equations
describing a sequence of dipole black rings and KK bubbles.

The functions $W$, ${\cal Y }$ and $\lambda$ are regular everywhere provided parameters $k_1$ and $k_2$  lie on any of the bubble rods

\begin{eqnarray}
a_{2m-1} <k_2 <k_1 <a_{2m},
\end{eqnarray}
where $m=1,2,...,N/2$, and $\alpha$ and $\beta$ satisfy

\begin{eqnarray}
&&\alpha^2 = \prod^{2m-1}_{i=1}(k_1-a_i)^{(-1)^{i+1}} \prod^{N}_{j=2m}(a_j-k_1)^{(-1)^{j+1}} ,   \\ \nonumber   \\
&&\beta^2 = \prod^{2m-1}_{i=1}(k_2-a_i)^{(-1)^{i}} \prod^{N}_{j=2m}(a_j-k_2)^{(-1)^{j}} .
\end{eqnarray}

Conical singularities are avoided satisfying $p$ balance conditions on each of the bubble rods

\begin{eqnarray}
(\Delta \phi)_{Rod[a_{2s-1},a_{2s}]} = L, \;\; s=1,2,..., N/2.
\end{eqnarray}
where

\begin{eqnarray}
&&(\Delta \phi)_{Rod[a_{2s-1},a_{2s}]} = 2\pi \lim_{\rho\to 0}\sqrt{{\rho^2 g_{\rho\rho}\over g_{\phi\phi}}}= \left( {{\cal Y}\over W}\right)^{3/2}_{Rod[a_{2s-1},a_{2s}]}
(\Delta \phi)^{E}_{Rod[a_{2s-1},a_{2s}]}   \nonumber \\ \nonumber \\
&&\left({{\cal Y}\over W}\right)_{Rod[a_{2s-1},a_{2s}]}= \left[ 1 + \alpha\beta \prod^{2s-1}_{i=1} \left({k_1-a_i\over k_2-a_i }\right)^{(-1)^{i}}
   \over
1 + \alpha\beta \right]^2 \prod^{2s-1}_{i=1} \left({k_2-a_i\over k_1-a_i } \right)^{(-1)^{i}} \nonumber
\end{eqnarray}
$(\Delta \phi)^{E}$ is the period for the seed solution given by

\begin{equation}
 (\Delta \phi)^E_{Rod[a_{2s-1},a_{2s}]} =
4\pi  (a_N-a_1) \prod_{i=2}^{2s-1} \prod_{j=2s}^{N-1}
(a_j-a_i)^{(-1)^{i+j+1}} \prod_{i=2}^{2s-1}
[\sqrt{a_N-a_i}]^{(-1)^{i+1}} \prod_{i=2s}^{N-1}
[\sqrt{a_i-a_1}]^{(-1)^{i}} .\nonumber
\end{equation}

\paragraph{}The general expressions for the dipole charge and dipole potential characterizing the s-th black ring $(m < s)$ are respectively

\begin{eqnarray}
 Q_{s}&=& \frac{L\sqrt{3}\beta \Delta k \Delta_{s}(a_{2s} - k_2)^{-1}\prod^{N}_{i=2s+2}(a_i - k_2)^{(-1)^{i+1}} \left[ 1 + \alpha\beta\prod^N_{i=2s+1}\left(\frac{a_i - k_1}{a_i - k_2}\right)^{(-1)^i}\right]}{2\pi \left[1  + \alpha\beta\prod^N_{i=2s}\left(\frac{a_i - k_1}{a_i - k_2}\right)^{(-1)^i}\right] \left[ 1 + \alpha\beta\prod^N_{i=2s+2}\left(\frac{a_i - k_1}{a_i - k_2}\right)^{(-1)^i}\right]}, \nonumber \\  \\
{\cal B}_{s} &=& -\frac{2\sqrt{3}\alpha\Delta k \prod^{N}_{i=2s+1}(a_i - k_1)^{(-1)^{i}}}{ \left[ 1 + \alpha\beta\prod^N_{i=2s+1}\left(\frac{a_i - k_1}{a_i - k_2}\right)^{(-1)^i}\right]} \nonumber
\end{eqnarray}
where $\Delta_{s}$ denotes the horizon rod length $a_{2s+1} - a_{2s}$.

\paragraph{}The general expressions for the dipole charge and dipole potential characterizing the s-th black ring $(m > s)$ are respectively

\begin{eqnarray}
 Q_{s}&=& - \frac{L\sqrt{3}\alpha\Delta k \Delta_{s}(k_1 - a_{2s+1})^{-1}\prod^{2s-1}_{i=1}(k_1 - a_i)^{(-1)^{i}} \left[ 1 + \alpha\beta\prod^{2s}_{i=1}\left(\frac{a_i - k_1}{a_i - k_2}\right)^{(-1)^i}\right]}{2\pi \left[1  + \alpha\beta\prod^{2s-1}_{i=1}\left(\frac{a_i - k_1}{a_i - k_2}\right)^{(-1)^i}\right] \left[ 1 + \alpha\beta\prod^{2s+1}_{i=1}\left(\frac{a_i - k_1}{a_i - k_2}\right)^{(-1)^i}\right]}, \nonumber \\
{\cal B}_{s} &=& \frac{2\sqrt{3}\beta\Delta k \prod^{2s}_{i=1}(k_2-a_i)^{(-1)^{i+1}}}{ \left[ 1 + \alpha\beta\prod^{2s}_{i=1}\left(\frac{a_i - k_1}{a_i - k_2}\right)^{(-1)^i}\right]}
\end{eqnarray}
where $\Delta_{s}$ denotes the horizon rod length $a_{2s+1} - a_{2s}$.

\paragraph{}Similarly, we find for the temperature and entropy of the s-th black ring

\begin{eqnarray}
T_s &=& {\cal Y}^{-3/2}_{{{\cal H}_s}}T_s^E \\ \nonumber
S_s &=&  {\cal Y}^{3/2}_{{{\cal H}_s}} S_s^E \left(\frac{L}{L^E}\right),
\end{eqnarray}
where

\begin{equation}
{\cal Y}_{{\cal H}_s} = \left[\frac{1+\alpha\beta\prod^N_{i=2s+1}\left( \frac{a_i -k_1}{a_i - k_2} \right)^{(-1)^i}}{1 + \alpha\beta}\right]^2 \prod^N_{i=2s+1}\left( \frac{a_i -k_2}{a_i - k_1} \right)^{(-1)^i} \nonumber
\end{equation}
and

\begin{eqnarray}
T_s^E &=& \frac{1}{4\pi} (a_N-a_1)^{-1}
\prod_{i=2}^{2s} \prod_{j=2s+1}^{N-1} (a_j-a_i)^{(-1)^{i+j}}
\prod_{i=2}^{2s} [\sqrt{a_N-a_i}]^{(-1)^{i}}
\prod_{i=2s+1}^{N-1} [\sqrt{a_i-a_1}]^{(-1)^{i+1}}, \nonumber \\
 \\
S_s^E &=& \frac{L^E}{4T_s^E}(a_{2s+1}-a_{2s}),
\end{eqnarray}
are the temperature and entropy corresponding to the s-th event horizon of the seed solution and the metric function ${\cal Y}$ is evaluated on the horizon rod $a_{2s}<z<a_{2s+1}$.

The magnetic fluxes are given by

\begin{eqnarray}
&&\Psi^{+}= L {\sqrt{3} \Delta k \beta (a_N-k_2)^{-1}\over \left[1 + \alpha\beta {a_N-k_1\over a_N-k_2} \right]  }, \\
\nonumber \\
&&\Psi^{-}=- L  {\sqrt{3} \Delta k \alpha (k_1-a_1)^{-1}\over \left[1 + \alpha\beta {k_2-a_1\over k_1-a_1} \right]  }.
\end{eqnarray}

\end{document}